\documentclass[10pt,conference]{IEEEtran}
\usepackage{cite}
\usepackage{amsmath,amssymb,amsfonts}
\usepackage{mathtools}
\usepackage{algorithmic}
\usepackage{graphicx}
\usepackage{booktabs}
\usepackage{textcomp}
\usepackage{xcolor}
\usepackage{physics}

\usepackage[normalem]{ulem}    
\usepackage[hidelinks]{hyperref}

\newcommand{\emaillink}[1]{\href{#1}{#1}}

\begin{document}

\title{X-parameter based design and simulation of Josephson traveling-wave parametric amplifiers for quantum computing applications
}

\author{\IEEEauthorblockN{Kaidong Peng}
\IEEEauthorblockA{\textit{Department of Electrical Engineering and Computer Science} \\
\textit{Massachusetts Institute of Technology}\\
Cambridge, United States \\
\emaillink{kdpeng@mit.edu}}
\and
\IEEEauthorblockN{Rick Poore}
\IEEEauthorblockA{\textit{PathWave Software and Solutions} \\
\textit{Keysight Technologies, Inc.}\\
Calabasas, United States \\
\emaillink{rick\_poore@keysight.com}}
\and
\IEEEauthorblockN{Philip Krantz}
\IEEEauthorblockA{\textit{Quantum Engineering Solutions} \\
\textit{Keysight Technologies Sweden AB}\\
Gothenburg, Sweden \\
\emaillink{philip.krantz@keysight.com}}
\and
\IEEEauthorblockN{David E. Root}
\IEEEauthorblockA{\textit{Keysight Laboratories} \\
\textit{Keysight Technologies, Inc.}\\
Santa Rosa, United States \\
\emaillink{david\_root@keysight.com}}
\and
\IEEEauthorblockN{Kevin P. O'Brien}
\IEEEauthorblockA{\textit{Department of Electrical Engineering and Computer Science} \\
\textit{Massachusetts Institute of Technology}\\
Cambridge, United States \\
\emaillink{kpobrien@mit.edu}}
}

\maketitle

\begin{abstract}

We present an efficient, accurate, and comprehensive analysis framework for generic, multi-port nonlinear parametric circuits, in the presence of dissipation from lossy circuit components, based on “quantum-adapted” X-parameters. We apply this method to  Josephson traveling-wave parametric amplifiers (JTWPAs) – a key component in superconducting and spin qubit quantum computing architectures – which are  challenging to model accurately due to their thousands of linear and nonlinear circuit components. 
X-parameters are generated from a harmonic balance solution of the classical nonlinear circuit and then mapped to the field ladder operator basis, so that the energy associated with each of the multiple interacting modes corresponds to photon occupancy, rather than classical power waves. Explicit relations for the quantum efficiency of a generic, multi-port, multi-frequency parametric circuit are presented and evaluated for two distinct JTWPA designs.  The gain and quantum efficiency are consistent with those obtained from Fourier analysis of time-domain solutions, but with enhanced accuracy, speed, and the ability to include real-world impairments, statistical variations, parasitic effects, and impedance mismatches (in- and out-of-band) seamlessly. The unified flow is implemented in Keysight’s PathWave Advanced Design System (ADS) and independently in an open-source simulation code, JosephsonCircuits.jl, from the MIT authors.

\end{abstract}

\begin{IEEEkeywords}
quantum-limited amplifiers, Josephson traveling-wave parametric amplifier (JTWPA), cryogenic parametric amplifiers, parametric devices, X-parameters, harmonic balance, qubit readout, quantum-mechanical loss model.
\end{IEEEkeywords}

\section{Introduction}
Nonlinear systems have abundant applications in both classical and quantum settings, ranging from frequency converters and mixers in classical circuits \cite{pozar_2011_microwave} to quantum-limited amplifiers \cite{yurke_observation_1989,macklin_a_2015} and transducers for quantum computing applications \cite{mirhosseini_superconducting_2020}. One of the fundamental cornerstones of building large-scale quantum information processors is access to high-fidelity and  scalable qubit readout \cite{DiVincenzo_2000}. The intrinsically low-power operation and frequency-domain multiplexed architecture of dispersive readout – commonly used to readout \textit{e.g.} superconducting\cite{Wallraff_2004,Kjaergaard_2020} and spin qubits \cite{Hornibrook_2014,Mi_2018,Landig_2018} – lays out requirements on signal-to-noise ratio, frequency bandwidth, as well as saturation power of the amplifier chain. To this end, the past decade has seen a rapid development of parametric amplifier technology used to pre-amplify the few-photon level (\textit{e.g.}$-150~$dBm) readout signals used to probe readout resonators operating at 10$~$mK to determine the qubit states\cite{krantz_a_2019}. Josephson parametric amplifiers (JPAs) \cite{yurke_observation_1989, castellanos-beltran_widely_2007, yamamoto_flux_2008, bergeal_phase_2010} and traveling-wave parametric amplifiers utilizing either nonlinear kinetic inductance \cite{eom_a_2012, vissers_low_2016, malnou_three_2021} or Josephson nonlinearity \cite{obrien_resonant_2014,macklin_a_2015,white_traveling_2015,planat_photonic_2020,ranadive_kerr_2022, peng_floquet_2022} have achieved near-quantum limited noise performance and are now routinely used in superconducting circuit experiments and quantum sensing. JPAs have near-ideal intrinsic quantum efficiency and often consist of one or a few nonlinear resonators. They are compact in footprint and have well-understood behavior \cite{boutin_effect_2017,eichler_controlling_2014}.  Through impedance and nonlinearity engineering, their bandwidth \cite{mutus_strong_2014,roy_broadband_2015} and dynamic range \cite{frattini_optimizing_2018,naaman_high_2019,sivak_josephson_2020} can be improved.

Josephson traveling wave parametric amplifiers (JTWPAs) are attractive components given their broad bandwidth (e.g. $3$ GHz), high gain ($20$ dB), high saturation power ($-100$ dBm), and noise performance approaching the fundamental limit allowed by quantum mechanics \cite{obrien_resonant_2014,macklin_a_2015,planat_photonic_2020, peng_floquet_2022}. However, despite the suitable characteristics for qubit readout applications, JTWPAs typically consist of thousands of Josephson junctions and thus exhibit more complicated dynamics than JPAs. The design and simulation of JTWPAs with desired performance still constitute a practical bottleneck for many quantum engineers. While simple, analytical two-mode wave-equation models \cite{yaakobi_parametric_2013,obrien_resonant_2014,grimsmo_squeezing_2017,houde_loss_2019} are commonly used for device design and modeling, they fall short in predicting realistic multi-mode gain dynamics\cite{dixon_capturing_2020} and quantum efficiency \cite{macklin_a_2015, peng_floquet_2022} and in capturing non-ideal behaviors observed in experiments, such as the cutoff frequency and photonic bandgaps that originate from the lumped-element nature of the JTWPA design \cite{planat_photonic_2020}, as well as parameter variability associated with fabrication uncertainties \cite{tolpygo_fabrication_2015,Kreikebaum_improving_2020,osman_simplified_2021,wan_fabrication_2021}. A harmonic balance method analysis of the signal gain in kinetic inductance traveling-wave parametric amplifiers was performed in \cite{sweetnam_simulating_2022}; however, for quantum efficiency calculations, the full X-parameter sensitivity matrix must be calculated including for inputs at all of the idler frequencies generated in the parametric processes.

In this work, we present an efficient and accurate circuit-level framework for analyzing the quantum statistics of generic multi-port parametric quantum circuits. This framework is directly applicable to current-pumped JTWPAs exemplified here and is embedded into both the X-parameter analysis flow of Keysight’s Pathwave Advanced Design System (ADS) and independently in an open-source simulation code, JosephsonCircuits.jl \cite{josephsoncircuits_jl}, from the MIT authors. To the best of author's knowledge, this work presents the first quantum efficiency calculation of a circuit quantum electrodynamics parametric device, including loss, using a commercial circuit simulator. The analysis framework is applicable to all harmonic balance solvers. In addition, the unified design, simulation, and analysis flow based on X-parameters can also be applied to model applications requiring multiple pump tone excitations, such as the generation of broadband squeezed vacuum as quantum resource \cite{qiu_broadband_2022}, and can be used to integrate multiple nonlinear components and model the overall system measurement efficiency of a qubit readout chain.

\begin{figure}[t]
\centerline{\includegraphics{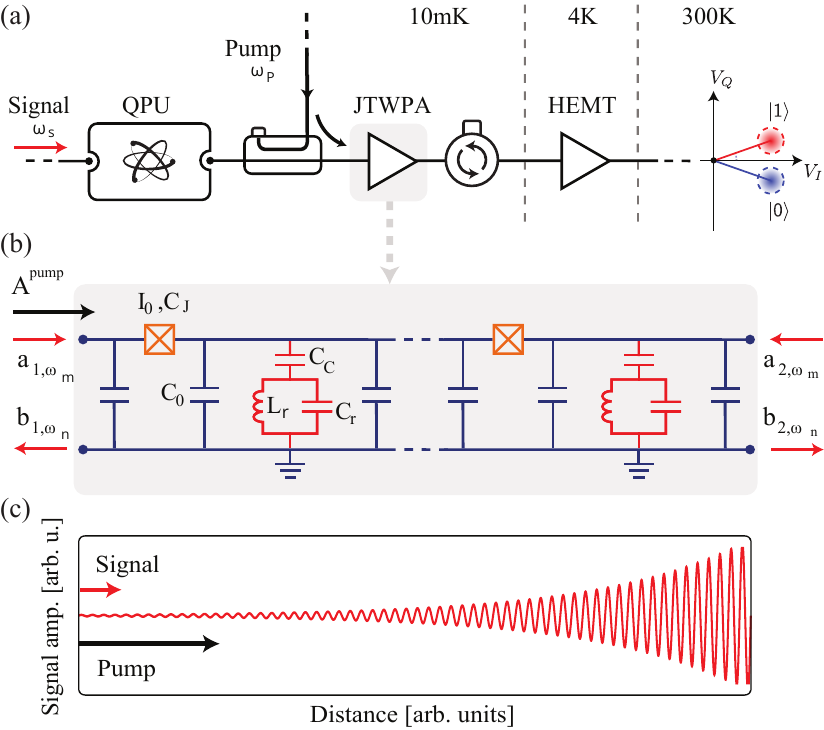}}
\caption{(a) Simplified schematic of the readout circuitry for a quantum processor (QPU) probed by a weak signal that gets amplified by a JTWPA. (b) Circuit schematic of JTWPA as a lumped-element transmission line built up from unit cells of Josephson junctions (orange boxes), shunt capacitors, as well as phase-matching resonators (red). Note that a realistic device has several Josephson junctions between each  phase-matching resonator and dielectric losses which can be accounted for in simulations using a complex capacitance. (c) Cartoon of amplified signal along the length of the JTWPA.}
\label{fig:fig1}
\end{figure}

Fig. \ref{fig:fig1}(a) shows a diagram of a typical readout-signal path passed through a quantum processor unit (QPU). The signal is then amplified using a JTWPA at the output of the quantum processor at the 10$~$mK-stage of a cryostat, before the signal reaches the high-electron mobility transistor (HEMT) amplifier at the 4$~$K-stage and eventually is down-converted and digitally processed using classical room-temperature electronics.

\begin{figure}[t]
\centerline{\includegraphics[width=\linewidth]{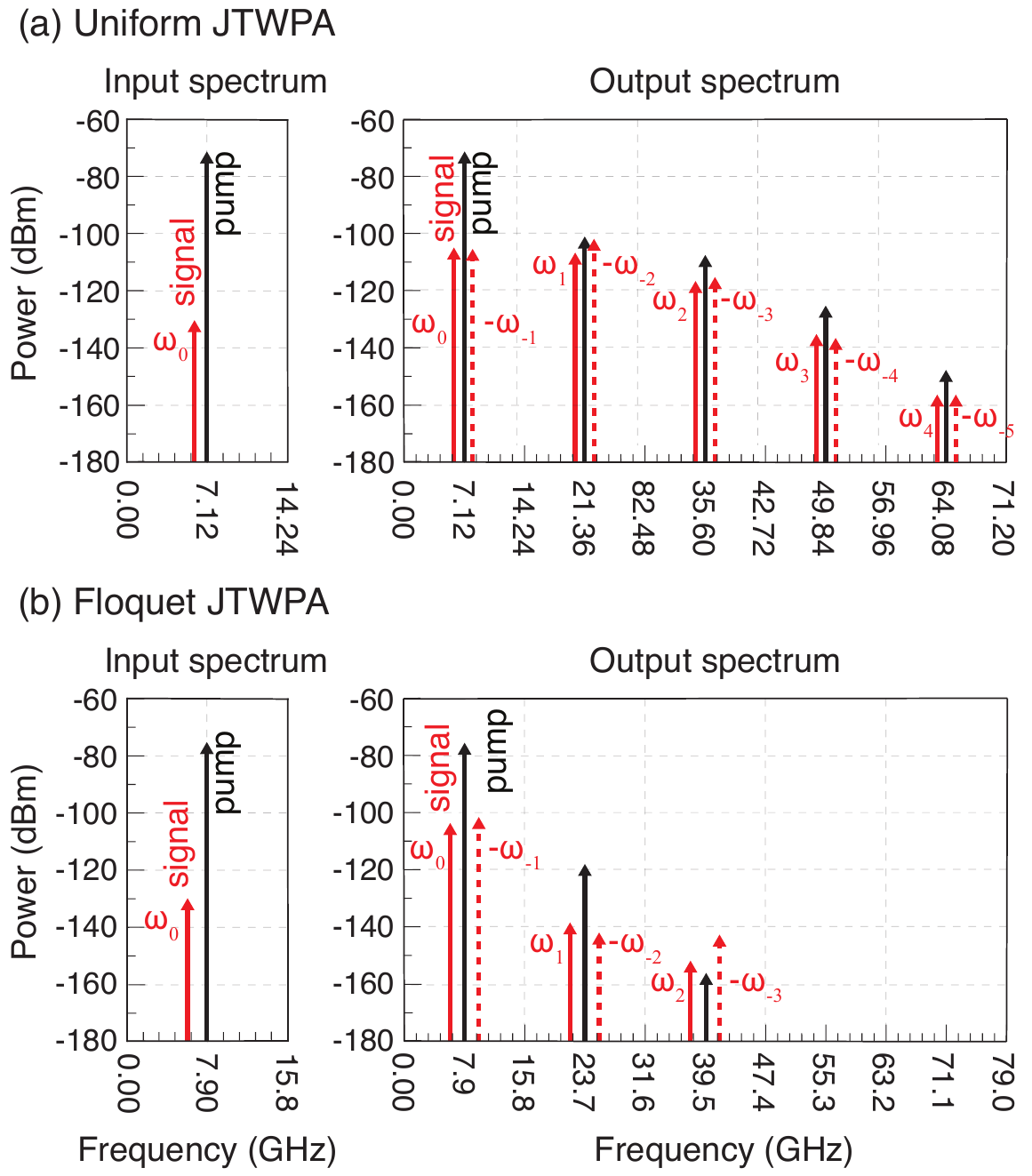}}
\caption{Single-sided input and output spectra for (a) a uniform JTWPA and (b) a Floquet JTWPA respectively, including the pumps and their harmonics. Frequencies are defined according to \eqref{eq:freqdefn}. Negative frequencies (dashed lines) from the double-sided representation are shown as upper sidebands with respect to the pump harmonics. The input signal power was chosen to be $-130~$dBm such  that the higher order sidebands of the Floquet JTWPA are visible on the spectrum, and the amplification results in negligible pump depletion.}
\label{fig:fig2}
\end{figure}

We model the JTWPA as a two-port traveling wave amplifier that converts energy from a large pump tone at angular frequency $\omega_P$ to a much smaller in power signal at a frequency $\omega_s$, see Fig. \ref{fig:fig1}(a), which is then amplified. The nonlinearity required for parametric amplification originates from the Josephson junctions (symbols marked with orange “x” in boxes in Fig. \ref{fig:fig1}(b)). Josephson junctions operate based on quantum tunneling of Cooper-pairs through thin barriers, and in this context can be treated as lossless nonlinear inductors. Lack of dissipation is critical for low added noise and to avoid heating the milli-Kelvin environment needed for high-fidelity qubit operation. As the pump and its harmonics interact with the signal, many intermodulation terms are created and propagate further along the structure. Typical input and output spectra are shown in Fig. \ref{fig:fig2}.

Design challenges for JTWPAs include engineering the circuit parameters to simultaneously achieve high gain, broad bandwidth, and high quantum efficiency in the presence of realistic fabrication constraints. The parametric amplification process must be phase matched by engineering the nonlinearity or the dispersion by, e.g. creating a stopband around the pump frequency using phase-matching resonators \cite{obrien_resonant_2014,macklin_a_2015,white_traveling_2015} (see red subcircuits in Fig. \ref{fig:fig1}(b)) or through periodic loading\cite{planat_photonic_2020}. A noteworthy recent development has been the “Floquet” design, whose junction critical current is spatially varied to realize an effective Gaussian nonlinearity profile along the structure. Consequently, the Floquet JTWPA is able to coherently recover information that would otherwise be lost to other sideband modes by adiabatically mode-matching the collective Floquet modes of the nonlinear system to the single frequency modes of the input and output ports, resulting in a quantum efficiency approaching the ideal limit \cite{peng_floquet_2022}. The designs of the Floquet-mode JTWPA in \cite{peng_floquet_2022} and the widely used uniform JTWPA in\cite{macklin_a_2015} are analyzed using the new methods presented in this work. Their circuit parameters are summarized in Table \ref{tab:twpaparameters}. Manufacturing tolerances and other real-world impairments create additional challenges, requiring robust, practical, and flexible modeling and design capabilities to realize optimal performance as part of the quantum computer readout chain.

\begin{table}[t]
\begin{center}
\caption{Design and operation parameters for the uniform and Floquet JTWPA designs. The pump current $I_P$ represents the peak current for a current source with a 50$\Omega$ source impedance. Approximately half of the pump current enters the JTWPA. $P_P$ denotes the available pump power.}
\begin{tabular}{c|ccccc}
\toprule
\textbf{Design} & JJ & $\omega_{P}/2\pi$ & I$_P$ & P$_P$ & JJ I$_c$ \\
\textbf{ } & Count & (GHz) & ($\mu$A) & (dBm) & ($\mu$A) \\ \midrule
\textbf{Uniform} & 2047   & 7.12  & 3.70  & -70.68      & 3.4        \\ 
\textbf{Floquet} & 3998   & 7.90  & 4.40  & -69.17     & 3.50-21.21 \\ 
\bottomrule
\end{tabular}
\label{tab:twpaparameters}
\end{center}
\end{table}

\section{Classical and Quantum X-parameter Formulations of JTWPA}
\subsection{Classical X-parameters}
Analytical two-mode continuous models \cite{yaakobi_parametric_2013,obrien_resonant_2014,grimsmo_squeezing_2017,houde_loss_2019} consider only the small signal injected at frequency $\omega_s$ and the ``image frequency" or primary \textit{idler} at frequency $-2\omega_P + \omega_s$. Thus, they would always predict an ideal quantum efficiency for a lossless JTWPA, inconsistent with experimental observations \cite{macklin_a_2015}. To accurately model the system dynamics and the quantum efficiency of JTWPAs, we go beyond the two-mode model and choose a larger number of modes $m$ to account for the effects of the interacting sidebands modes \cite{mcKinstrie_quantum_2005,chaudhuri_simulation_2015,dixon_capturing_2020, peng_floquet_2022}. We find that an accurate computation of quantum efficiency for typical JTWPA designs requires $m\geq6$.  For arbitrary positive signal and pump frequencies $\omega_s$ and $\omega_P$ respectively, a discrete ordered set of frequencies – that can be positive or negative – defines the $k$-th signal/idler frequency according to

\begin{equation}
    \omega_{k} \equiv 2 k \omega_{P}+\omega_{s},
\label{eq:freqdefn}
\end{equation}

\noindent  for $k \in\{-\lfloor m / 2\rfloor, \ldots,\lfloor(m-1) / 2\rfloor\}$, where $\lfloor.\rfloor$ is the floor function, and $\omega_0=\omega_s$. The set $\{\omega_k\}$ defined by \eqref{eq:freqdefn} follows from the cosine potential energy of unbiased Josephson junctions \cite{mccumber_effect_1968} and the linearization with respect to the signal and generated idlers. Equation \eqref{eq:freqdefn} defines a double-sided intermodulation spectrum corresponding to first-order mixing of the signal with the odd harmonics of the pump frequency.

We work in conventional power-wave variables defined by the linear combinations of complex voltages and currents at each port according to $a_n = (V_n + Z_0I_n)/(2\sqrt{Z_0})$ and $b_n = (V_n - Z_0I_n)/(2\sqrt{Z_0})$, with $Z_0$ being a fixed characteristic reference impedance taken conventionally to be $50~\Omega$. The  relationship among incident waves $a_{q,\omega_k}$ at port $q$ at frequency $\omega_k$ and scattered waves $b_{p,\omega_j}$ at port $p$ at frequency $\omega_j$ is given by

\begin{equation}
    b_{p, \omega_{j}}=\sum_{\mathclap{  \substack{q=1,2 \\ k=-\lfloor m / 2\rfloor,\dots,\lfloor(m-1) / 2\rfloor }}} ~X_{p, \omega_{j} ; q, \omega_{k}} a_{q, \omega_{k}},
\label{eq:classicalxp}
\end{equation}

\noindent where $p \in \{1,2\}$ represents the two ports and $j \in \{- \lfloor m /2 \rfloor, \lfloor (m-1) /2 \rfloor\}$ is the mode index. Moreover, the coefficients in  \eqref{eq:classicalxp} can be identified as the elements of the  double-sided X-parameter sensitivity matrix of dimension $2m\times 2m$  \cite{biernacki_circuit_2017} for incommensurate signal and pump frequencies, expressible as

\begin{equation}
    X_{p, \omega_{j} ; q, \omega_{k}} \equiv \lim _{a_{q, \omega_{k}} \rightarrow 0} \frac{b_{p, \omega_{j}}\left(\left|A^{\text {pump }}\right|\right)}{a_{q, \omega_{k}}}.
\label{eq:X}
\end{equation}

The X-parameters in \eqref{eq:classicalxp} and \eqref{eq:X}  are nonlinear functions of the magnitude of the pump amplitude. The X-parameters are computed, not from \eqref{eq:X}, but rather by the simulator as part of the harmonic balance solution of the system circuit equations in the frequency domain, utilizing the spectral Jacobian computed from the well-defined partial derivatives of the constitutive relations of the circuit elements (Josephson junctions and linear capacitors and inductors) of the JTWPA. The resulting X-parameters are the wave-variable representation of the classical conversion matrix formalism usually expressed in admittance or impedance representation \cite{maas2003nonlinear}.

The full nonlinear X-parameter formalism, including the measurement science and the nonlinear mathematical model implemented in the ADS simulator, are much more general than the limited discussion of the X-parameter sensitivity matrix analysis that is the main focus of this work. The X-parameter sensitivity matrix introduced in \eqref{eq:classicalxp} and  \eqref{eq:X} is just the spectrally linearized portion of a more complete X-parameter model that includes the contributions to the scattered waves from each port due to the pump and its harmonics in addition to the terms linear in the signal phasor and created idlers. Together, these terms account for all the spectral components shown in Fig. \ref{fig:fig2}.  More generally and most importantly, X-parameter models build into the mathematical representation the required phase relationships among commensurately related signals, such as the pump harmonics and any reflections thereof, as required by the underlying physical constraint of time-invariance of the component. For example, the phasor at port $p$ at the $n$-th pump harmonic can be shown to have the general X-parameter form $X^F_{p,n\omega_p}\left(\abs{A_p}\right)e^{ni\theta(A_p)}$. These are the Fourier coefficients of waves that have, by construction, the required phase dependence to ensure that when the input signal is delayed by any amount the same output response is obtained as previously, but now delayed by precisely the same interval as the input delay.  The time-invariant constraint on the mathematical model is not satisfied for a general complex-valued nonlinear function of the complex input phasor $A_p$. X-parameter models identified from measurements require special nonlinear calibration and a cross-frequency phase calibration to identify the time-invariant intrinsic device characteristics. The co-variant mathematical formulation of the model terms guarantees that any simulations with the model will behave correctly for the entire equivalence class of signals delayed by an arbitrary time.  In some cases, such as a JPA, where the signal may be incident at a harmonic of the pump, the explicit phase-dependence of the amplification is properly taken into account by the X-parameter model \cite{root_x_2013}.

\subsection{From Classical to Quantum Properties}

The above analysis applies to the linearized dynamics of a nonlinear classical circuit driven by a strong, periodically time-varying pump. The linearized power-wave solutions can then be mapped into the quantum field ladder operator basis \cite{peng_floquet_2022} to properly calculate the quantum statistics of linear amplifiers \cite{caves_quantum_1982}. This is accomplished by defining quantum X-parameters according to \eqref{eq:qxparamter}. The transformation from the classical power wave basis to the quantum ladder operator basis \cite{roy_introduction_2016} can be justified rigorously by deriving the input-output relations in the field ladder operator basis from the JTWPA system circuit Lagrangian and the corresponding Heisenberg equations of motions \cite{peng_floquet_2022,josephsoncircuits_jl}. Alternatively, this transformation can be understood classically as a consequence of the Manley-Rowe relations which are an expression of energy conservation \cite{manley_some_1956}.

\begin{equation}
    x_{p, \omega_{j} ; q, \omega_{k}} \equiv X_{p, \omega_{j} ; q, \omega_{k}} \sqrt{\left|\frac{\omega_{k}}{\omega_{j}}\right|}. \label{eq:qxparamter}
\end{equation}

Equation \eqref{eq:commrelation} describes the additional constraint on the quantum X-parameters to preserve the underlying quantum commutation relations of bosonic operators \cite{haus_quantum_1962,caves_quantum_1982,clerk_introduction_2010,bergeal_analog_2010,peng_floquet_2022}. The quantum X-parameters numerically evaluated from the simulator have been checked to satisfy \eqref{eq:commrelation} within numerical precision, supporting the correctness of our framework.

\begin{equation}
    \sum_{\mathclap{\substack{q=1,2 \\ k=-\lfloor m / 2\rfloor, \dots,\lfloor(m-1) / 2\rfloor}}} ~\operatorname{sgn}\left(\omega_{k}\right)\left|x_{p, \omega_{j} ; q, \omega_{k}}\right|^{2}=\operatorname{sgn}\left(\omega_{j}\right). \label{eq:commrelation}
\end{equation}

\subsection{Quantum Efficiency from Quantum X-parameters}
Quantum efficiency (QE) is defined as the ratio of the output to input signal-to-noise ratio (SNR) \cite{boutin_effect_2017,peng_floquet_2022}. In the case when each of the input modes has a minimum vacuum noise of half a quantum, QE for a pair of input and output modes, $a_{q,\omega_l}$ and $b_{p,\omega_n}$, represents the ratio of the noise in the output mode $b_{p,\omega_n}$ originating from the input mode $a_{q,\omega_l}$ to the sum of noise from all input modes. A closed-form expression for the quantum efficiency in terms of the quantum X-parameters is given by

\begin{equation}
    \mathrm{QE}_{p, \omega_{n} ; q, \omega_{l}}=\frac{\left|x_{p, \omega_{n} ; q, \omega_{l}}\right|^{2}}{\sum\limits_{\mathclap{\substack{q'=1,2 \\ k=-\lfloor m / 2\rfloor,\dots,\lfloor(m-1) / 2 \rfloor}} }~\left|x_{p, \omega_{n} ; q^{\prime}, \omega_{k}}\right|^{2}}, \label{eq:qeexpr}
\end{equation}

\noindent which is a matrix, analogous to the noise figure matrix defined in \cite{mcKinstrie_quantum_2005}. Quantities of particular interest for transmission-mode quantum-limited amplifiers are $\mathrm{QE}_{2,\omega_0,1,\omega_0}$ and $\mathrm{QE}_{2,\omega_{-1},1,\omega_0}$ that define the quantum efficiency of the signal and the primary idler, respectively, for inputs at port 1 and outputs at port 2, including interactions with all $m$ modes. Other entries in the QE matrix, such as $\mathrm{QE}_{2,\omega_1,1,\omega_0}$ and $\mathrm{QE}_{2,\omega_{-2},1,\omega_{-1}}$ define the quantum efficiency of devices functioning as frequency converters. For the uniform JTWPA design considered here, the number of modes needs to be at least six before the results converge to a level of $10^{-3}$, confirming the limitation of two-mode models.

\subsection{Quantum Efficiency with Dissipation}

The impedance of a capacitor of capacitance $C_d$ with dielectric loss tangent $\tan \delta$ can be modeled as an effective series resistance plus a standard capacitor

\begin{equation}
    Z_d(\omega) = \frac{\tan \delta}{\omega C_d} + \frac{1}{j \omega C_d}
\end{equation}

\noindent or 
$Y_d(\omega) = j \omega C_d / \left( 1 + j \tan \delta \right)$
when expressed as an admittance, which is more efficient for simulation. The loss in the capacitor produces noise, which can be expressed as a noise current spectral density 
$\left< i_n^2 \right> (\omega)$
generated in parallel with the admittance $Y$

\begin{equation}
    \left< i_n^2 \right> (\omega) = 
    4 \frac{\hbar \omega \coth{ \left( \hbar \omega / 2 k T \right) }}{2} \operatorname{Re} \left[ Y(\omega) \right].
\end{equation}

\noindent At non-cryogenic temepratures and typical microwave frequencies it simplifies to the classical $4 k T \operatorname{Re} \left[ Y(\omega) \right]$; at $T=0$, this simplifies to $4 \left( \hbar \omega / 2 \right) \operatorname{Re} \left[ Y(\omega) \right]$.

The noise power delivered to port $p$ at frequency $\omega_n$, due to the noise generated by all of the components in the system, is calculated using 
\begin{equation}
    P^{N,dut}_{p, \omega_n} = 
    \mathop{ \sum}\limits_{k}
    \mathop{ \sum}\limits_{d} 
    Z_{d,k;p,n} \left< i^2_{n:d,k} \right> Z^\dagger_{d,k;p,n} / \operatorname{Re} \left( Z_{p,n} \right)
\end{equation}
    
\noindent where 

\begin{itemize}
\item $k=-\lfloor m / 2\rfloor,\dots,\lfloor(m-1) / 2 \rfloor$;
\item $d$ is the set of all noise generating devices;
\item $Z_{d,m;p,n} = \partial v_{p,n} / \partial i_{d,m}$ denotes the transimpedance from device $d$'s current at frequency index $m$ to the port $p$ at frequency index $n$;
\item $\left< i^2_{n:d,k} \right>$ is the noise correlation matrix due to device $d$ at frequency index $k$;
\item $Z_{p,n}$ is the impedance of port $p$ at frequency index $n$.
\end{itemize}

\noindent The available noise power from input port $q$ and frequency $\omega_l$ is given by
\begin{equation}
    P^{N,in}_{q, \omega_l} 
    = \hbar \omega_l \coth{ \left( \hbar \omega_l / 2 k T \right)} / 2.
\end{equation}

\noindent The ratio of these two noise values, after frequency scaling to photon number units, represents the fraction of added noise power to input noise power

\begin{equation}
    N_{p, \omega_{n}} = 
    \frac{P^{N,dut}_{p, \omega_n} / \abs{\hbar\omega_n}}
         {P^{N,in}_{q, \omega_l} / \abs{\hbar\omega_l}}.
\end{equation}

With a typical electrical length of tens of wavelengths, realistic JTWPA devices often have non-negligible propagation loss due to two-level-system defects in the dielectrics of the capacitors. Therefore, the quantum efficiency calculation at the presence of loss has to properly account for the excess noise introduced to the device output associated with dissipation, as dictated by the fluctuation-dissipation theorem \cite{Kubo_1966}. The quantum efficiency of a JTWPA with dissipation can be expressed by adding the noise ratio $N_{p, \omega_{n}}$ to the denominator of (\ref{eq:qeexpr})

\begin{equation}
    \mathrm{QE}_{p, \omega_{n} ; q, \omega_{l}}=\frac{\left|x_{p, \omega_{n} ; q, \omega_{l}}\right|^{2}}{\sum\limits_{\mathclap{\substack{q'=1,2 \\ k=-\lfloor m / 2\rfloor,\dots,\lfloor(m-1) / 2 \rfloor}} }~\left|x_{p, \omega_{n} ; q^{\prime}, \omega_{k}}\right|^{2} + N_{p, \omega_{n}} }. \label{eq:qelossexpr}
\end{equation}

The noise power response at the input and output ports can be calculated from the Green's function scattering matrices \cite{peng_floquet_2022} or using adjoint methods in classical circuit analysis \cite{rizzoli_general_1994,ngoya_on_2011}. Signal dissipation at each node can be understood as coherent scattering into a semi-infinite transmission line channel, and the associated noise is at the same time introduced into the system by a noise source radiating from the opposite side of the transmission line channel. Alternatively, the noise power response ratio can be conceptually understood as additional magnitude square quantum X-parameter terms if each internal node is also assigned a port in an extended X-parameter analysis. 

Analogous to \eqref{eq:commrelation}, \eqref{eq:losscommrelation} specifies the constraint on the noise power response terms and X-parameters to conserve bosonic commutation relations in the presence of loss.

\begin{equation}
\begin{aligned}
         \sum\limits_{\mathclap{\substack{k=-\lfloor m / 2\rfloor, \\
         \dots,\lfloor(m-1) / 2\rfloor}}}\operatorname{sgn}\left(\omega_{k}\right)\Bigg(\sum_{q=1,2}~\left|x_{p, \omega_{j} ; q, \omega_{k}}\right|^{2} + N_{p,\omega_n}  \Bigg)  = \operatorname{sgn}\left(\omega_{j}\right)
\end{aligned}
    \label{eq:losscommrelation}
\end{equation}

\section{Results}

\subsection{Gain vs Signal Frequency and Pump Power}
Each X-parameter matrix element is the power or conversion gain for one of the small incident signals. Specifically, the diagonal elements are power gains and the off-diagonals are conversion gains. For example, $\abs{X_{2,\omega_0;1,\omega_0}}^2$ gives the power gain of the signal at the output port relative to the input port, and $\abs{X_{2,\omega_{-1};1,\omega_0}}^2$ gives the conversion gain of the primary ($k=-1$) idler at the output port with respect to the signal incident at the input port.

\begin{figure}[t]
\centerline{\includegraphics[width=1\linewidth]{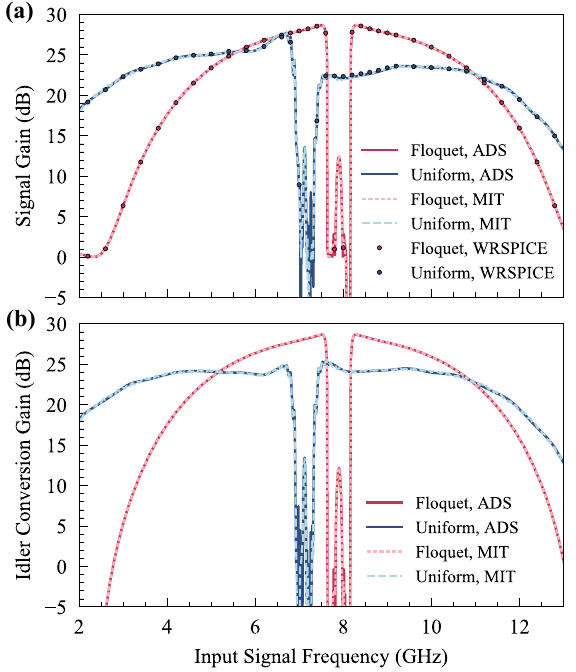}}
\caption{Simulated signal (a) and idler (b) gain versus signal frequency for two JTWPA designs, the uniform design (blue) and the Floquet design (red).}
\label{fig:fig3}
\end{figure}

Fig. \ref{fig:fig3} shows the gain of the signal and primary idler for the two JTWPA designs  \cite{macklin_a_2015,peng_floquet_2022} evaluated in this work. In both cases, the results obtained using frequency domain X-parameter analysis are consistent with nonlinear time-domain analysis using WRSPICE \cite{whiteley_josephson_1991} once the maximum phase change per step is reduced to 0.01 from the default of $\pi/5$ through the WRSPICE configuration parameter dphimax. On a desktop computer (AMD Ryzen 9 3950X 16-Core Processor), simulating gain and quantum efficiency for all 131 signal frequencies using JosephsonCircuits.jl with $m=10$ takes 3.9 and 2.5 seconds for the uniform and Floquet designs, respectively. In the Julia model, simulation time approximately doubles when dissipation is included. In WRSPICE, simulating the gain for a single frequency takes 13.3 and 8.2 seconds for the uniform and Floquet designs, respectively, and three simulations are required to extract the complex scattering parameters (pump only and pump + signal with two different phases). Simulating the required terms in the X parameter matrix to compute the quantum efficiency matrix ($2\cdot 2 \cdot m\cdot 131+1$ simulations) for 131 signal frequencies in WRSPICE would take an estimated 19.5 and 12.0 hours, respectively, which is order of 10,000 times slower than the Julia harmonic balance code. WRSPICE also is unable to model lossy capacitors. The ADS and MIT frequency domain analyses show excellent agreement. The MIT code, written specifically for these types of circuits, is naturally faster than a general-purpose microwave circuit simulator like ADS.

For periodic or quasi-periodic parametric circuits, harmonic balance (HB) is more accurate and faster than  time-domain solutions that integrate the ordinary differential equations until steady-state before performing Fourier analysis, with the linearization being approximated by subtracting the large-signal pump solution from a time-domain solution that includes small-signal tones as well.

\begin{figure}[t]
\centerline{\includegraphics[width=1\linewidth]{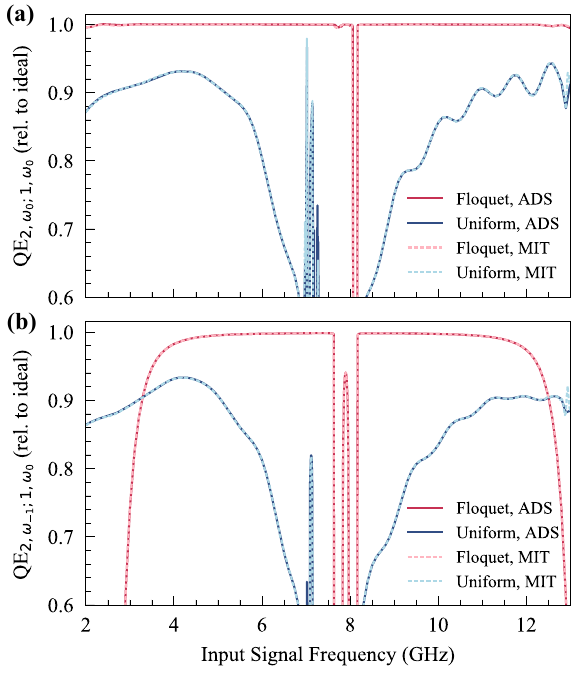}}
\caption{The simulated (a) signal and  (b) idler quantum efficiencies, QE$_{2,\omega_0, 1,\omega_0}$ and QE$_{2,\omega_{-1}, 1,\omega_0}$, respectively, of the uniform (blue) and Floquet JTWPA (red) designs normalized by those of an ideal quantum-limited amplifier at the same power gain $G$ specified by \eqref{eq:qeideal}.}
\label{fig:fig4}
\end{figure}

\subsection{Quantum Efficiency and the Number of Interacting Modes}

The signal quantum efficiency of the two considered JTWPA designs simulated with different number of modes $m$ at $\omega_0/(2\pi)=6\,$GHz is presented in Table \ref{tab:qevalues}. The QE values here are normalized by the corresponding ideal quantum efficiency at the same power gain $G$ to compare amplifier performance at similar but nonidentical gains \cite{boutin_effect_2017,peng_floquet_2022}:

\begin{equation}
\mathrm{QE}_{\mathrm{ideal}}(G) = \begin{cases}
    \frac{1}{2-1/G}, ~ &\abs{G} \geq1, \\
    1, ~ &\abs{G} <1.
\end{cases} \label{eq:qeideal}
\end{equation}

Consequently, ideal quantum-limited amplifiers have a normalized quantum efficiency of $100\%$ using this definition. For the uniform JTWPA design, the quantum efficiency normalized to the ideal converges to approximately $80.4\%$ at $m=6$, whereas the Floquet-mode JTWPA design converges to approximately $100.0\%$ at $m=4$. To the authors’ knowledge, this is the first quantum efficiency calculation of a circuit quantum electrodynamics device performed in a commercial circuit simulator.

\begin{table}[b]
\begin{center}
\caption{Simulated quantum efficiency vs number of modes $m$}
\begin{tabular}{c|ccccc}
\toprule
\textbf{Design} &  $m$=2 & $m$=4 & $m$=6 & $m$=8 & $m$=10\\ \midrule
\textbf{Uniform} & 0.999554  & 0.772218  & 0.804080  & 0.804028   & 0.804109\\
\textbf{Floquet} & 0.999998  & 0.999863  & 0.999857  & 0.999856 & 0.999856\\ \bottomrule
\end{tabular}\label{tab:qevalues}
\end{center}
\end{table}

Fig. \ref{fig:fig4} shows the signal and idler quantum efficiency QE$_{2,\omega_0, 1,\omega_0}$ and QE$_{2,\omega_{-1}, 1,\omega_0}$ from \eqref{eq:qeexpr}, normalized by the corresponding ideal quantum efficiency, as a function of the input signal frequency at $m=10$. Our framework predicts normalized signal quantum efficiency of  $81.35\%$ at $\omega_{0}/(2\pi)=5.984\,$ GHz for the uniform JTWPA design, consistent with the experimentally inferred value, factoring out dissipation, of $85\%$ in \cite{macklin_a_2015}. Our framework likewise yields a near-ideal quantum efficiency ($\sim\!99.9\%$) performance over the entire working bandwidth of the Floquet JTWPA design, corroborating the predictions of the wave-equation-based model in \cite{peng_floquet_2022} by also accounting for both the large number of interacting sideband modes and the discreteness of device design. The ability to accurately predict and differentiate the quantum noise performance of near-quantum-limited and quantum-limited amplifiers makes our framework an invaluable tool for the design and optimization of quantum parametric devices.

\subsection{Quantum Efficiency Calculation in the Presence of Loss}

\begin{figure}[t]
\centerline{\includegraphics[width=1\linewidth]{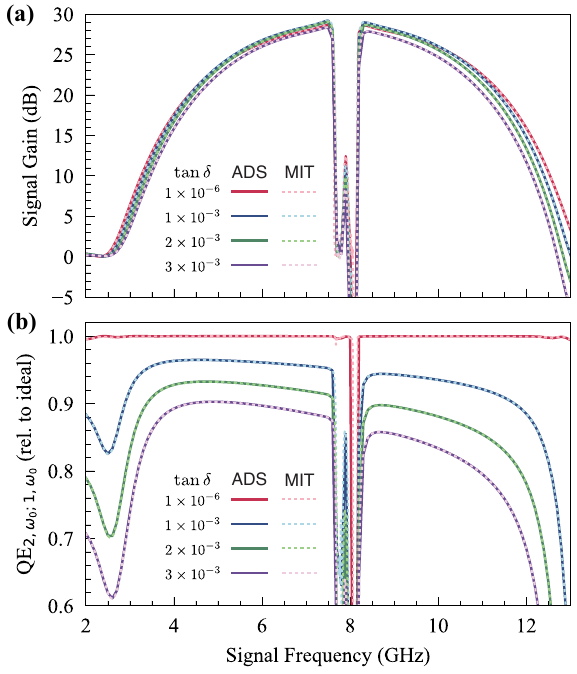}}
\caption{ The simulated (a) gain and (b) quantum efficiency spectra of the Floquet JTWPA design at several dielectric loss tangent values relevant to existing fabrication processes. The quantum efficiency is normalized by those of an ideal quantum-limited amplifier having the same gain ($1/(2 - 1/G)$ \cite{peng_floquet_2022} as in Fig. \ref{fig:fig4}.}
\label{fig:fig5}
\end{figure}

Fig. \ref{fig:fig5} shows the signal gain and quantum efficiency of the Floquet JTWPA design at several dielectric loss tangent values ranging from $10^{-6}$ to $3\cdot 10^{-3}$, pertinent to coplanar-waveguide capacitors and silicon dioxide or alumina parallel-plate capacitors \cite{macklin_a_2015,planat_photonic_2020}, respectively. We scale the input pump current in each trace by a factor of $(1 +125\tan\delta)$ to maintain a similar level of gain across different $\tan \delta$ value simulations for comparison. The quantum efficiency values predicted by ADS and the Julia model again show excellent agreement with each other, and the quantum efficiency decreases with increased dielectric loss tangent as expected. Different from the prediction in \cite{peng_floquet_2022} where the same input pump current and pump profile were assumed across different loss levels and resulted in significantly reduced signal gain, the Floquet TWPA has a predicted quantum efficiency close to $90\%$ even at a loss level $\tan\delta=3\times10^{-3}$ of silicon dioxide parallel-plate capacitors \cite{macklin_a_2015,tolpygo_fabrication_2015}. Furthermore, the quantum efficiency of the Floquet TWPA design at a certain loss level can be improved further by  optimizing its nonlinearity profile to balance the incoherent material loss and coherent sideband loss contributions, making the demonstration of a  near-term advantage of Floquet TWPA possible with current fabrication processes.

In addition, the bosonic commutation relations of the output modes at the presence of loss specified in \eqref{eq:losscommrelation} were numerically verified to be $\pm 1$ within numerical precision as required by quantum mechanics, suggesting the validity of both approaches as quantum-mechanically consistent models. 

\begin{figure}[bp]
\centerline{\includegraphics[width=1\linewidth]{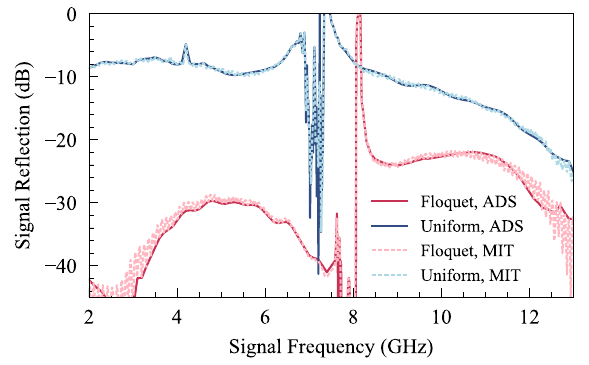}}
\caption{The signal $S_{11}$ (input match to 50$\,\Omega$ environment) of the uniform and Floquet JTWPA designs.}
\label{fig:fig6}
\end{figure}
\subsection{Mismatch Effects}

We can simulate the interaction of the JTWPA with neighboring components (e.g. cables to HEMT amplifiers at higher temperature stages of the measurement setup), out-of-band impedance effects, and the degree to which the input of the JTWPA is matched to the signal over the operating frequency range. An example is shown in Fig. \ref{fig:fig6}, where we plot the signal input mismatch of the chosen JTWPA designs to a $50\,\Omega$ environment as a function of signal frequency. We observe that despite having similar impedance and signal gain level, the uniform conventional JTWPA design results in significantly larger reflections compared to the Floquet JTWPA designs, further confirming the predictions in \cite{peng_floquet_2022} using a circuit model. The capability to quantitatively evaluate and engineer input and output matches provides valuable guidance in realizing applications such as on-chip integration of quantum-limited amplifiers.

\begin{figure}[tp]
\centerline{\includegraphics[width=1\linewidth]{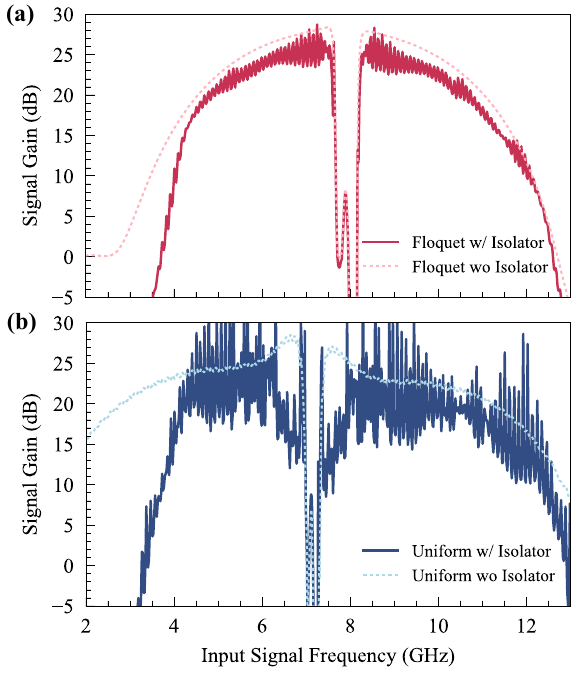}}
\caption{Simulated gain spectra of the (a) Floquet and (b) uniform JTWPA designs when placed in between a pair of 4-12 GHz commercial double-junction circulators, a common configuration in superconducting qubit measurement setup. A dielectric loss tangent of $\tan\delta=3\cdot10^{-3}$ is used for both TWPA designs, and the input pump current for the Floquet and uniform TWPA are scaled by $(1+125\tan\delta)$ and $(1+170\tan\delta)$, respectively, to maintain $>20\,$dB gain levels similar to those in Fig. \ref{fig:fig3}.  }
\label{fig:fig7}
\end{figure}

Impedance mismatch effects can be also straightforwardly accounted for and modeled in our framework. Significant out-of-band impedance mismatch occurs in a qubit readout chain as dictated by the Bode-Fano limit \cite{bode_network_1945,fano_theoretical_1950}, when in-band mismatch is optimized to maximize measurement efficiency and minimize qubit dephasing caused by undesirable back reflections. Nonlinear devices may be adversely affected by out-of-band impedance mismatch on sideband modes \cite{peng_floquet_2022}. In Fig. \ref{fig:fig7} we plot the predicted gain performance of the two JTWPA designs using a realistic dielectric loss tangent of $\tan\delta=3\cdot10^{-3}$ and under the realistic impedance environment resulted from a Quinstar commercial 4-12 GHz circulator (model QCY-G0401202AM) commonly used in superconducting qubit measurements using ADS. In ADS, the measured scattering parameters of the circulator considered are imported and used to directly specify the response of the created components. The JTWPA and circulators are simulated together to predict the performance in a realistic superconducting qubit measurement setup. Consistent with the predictions in \cite{peng_floquet_2022} using a simplified constant out-of-band impedance model, the uniform and Floquet TWPA designs show drastically different behaviors at the presence of out-of-band impedance caused by circulators. The uniform JTWPA shows instability and significantly increased gain ripples from parametric oscillations, qualitatively agreeing with the experimental observations that are not captured by models assuming a constant 50 $\Omega$ port impedance.

The gain profile of the Floquet JTWPA is less affected by the presence of the isolators than the gain profile of the uniform JTWPA. The small ripples are caused by both the large round trip gain ($\sim25$ dB) and the in-band impedance mismatch of the isolators. The slightly reduced bandwidth is caused by the 4-12 GHz working bandwidth of the isolators. Compared to the uniform TWPA designs, Floquet TWPA designs are much less sensitive to non-ideal out-of-band impedance mismatch due to more than an order of magnitude suppression of higher pump harmonics and sideband generation as seen in Fig. \ref{fig:fig2}. The ability of our analysis framework to model different in- and out-of-band impedance environments will be valuable for both modeling and designing the parametric devices under realistic experimental setups.

\subsection{Nonlinear Signal Effects}
\begin{figure}[tp]
\centerline{\includegraphics[width=1\linewidth]{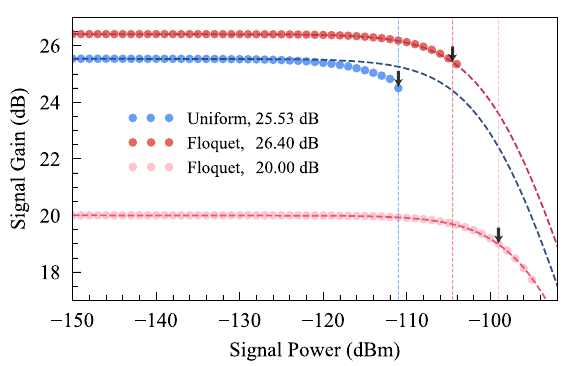}}
\caption{Gain compression and P$_{1\mathrm{dB}}$ of the two analyzed JTWPA designs. The dots are simulated using a full two-tone harmonic balance model in ADS, whereas the dashed lines are calculated using the analytical pump depletion formula in \cite{kylemark_semi_2006} (see main text). The P$_{1\mathrm{dB}}$ points are marked using arrows.}
\label{fig:fig8}
\end{figure}
The above treatment and results are based on an exact numerical linearization around the periodically time-varying solution of the JTWPA when excited by the large pump tone. We note that exact linearization is required for the standard quantum statistics analysis of linear amplifiers to hold \cite{caves_quantum_1982,clerk_introduction_2010}. When the signal power to be amplified becomes large enough there will arise compression and higher signal order intermodulation effects that need to be considered. Nonlinear signal effects are particularly detrimental in the context of frequency-multiplexed qubit readout, as they set an upper bound on the maximum number of qubits that can be read out simultaneously. These can be accounted for classically by replacing the linearized harmonic balance calculations with full two-tone harmonic balance analyses in which higher than first order in the signal and idler intermodulation products are simulated. A quantum mechanical analysis of these nonlinear signal effects is beyond the scope of this work.

Fig. \ref{fig:fig8} shows the compression characteristics of the signal gain versus input signal power for the uniform and Floquet JTWPA designs. Two signal gain compression curves are presented for the Floquet JTWPA at two different pump powers to highlight the dependence of the 1-dB gain compression point on the magnitude of the small signal gain. The gain compression points, $P_{1\mathrm{dB}}$,  are $-104.5$ dBm and $-99$ dBm at $26.40$ dB and $20.00$ dB gain, respectively. The simulated $P_{1\mathrm{dB}}$ of the uniform JTWPA design at $25.53$ dB gain is $-111$ dBm. We note that this value is smaller than the reported value in \cite{macklin_a_2015} of $-99\,$dBm due to differences in the operating pump configuration and small signal gain, the non-negligible material loss, and the non-ideal out-of-band impedance mismatch \cite{peng_floquet_2022} in the measured device.

We observe that the simulated signal gain compression values of the Floquet JTWPA design using the full two-tone harmonic balance simulations show excellent agreement with those predicted by the approximate analytical formula $G(P_s) = G_0 / (1 + 2G_0P_s/P_p)$ \cite{kylemark_semi_2006}, which captures the effect of pump depletion from the signal and the principal idler. This suggests that, just as the case for ideal two-mode parametric amplifiers, pump depletion is the dominant source for gain compression for the Floquet JTWPA designs, despite their more complicated circuit profiles. In contrast, the gain compression profile of the considered uniform JTWPA design deviates from that of the analytical pump depletion formula and leads to a lower $P_{1\mathrm{dB}}$ value. This may result from the additional pump power consumption due to the generation of higher pump harmonics and higher-order sideband modes by the uniform TWPA design, both of which are more than an order of magnitude larger than in the Floquet TWPA as shown in Fig. \ref{fig:fig2}.

\section{Conclusion}
A unified design and simulation methodology for generic multi-port, multi-frequency parametric circuits in the presence of loss based on quantum-adapted X-parameter analysis has been presented with example simulations for JTWPAs, a key amplifier technology widely used for superconducting qubit readout and quantum sensing. Each X-parameter provides a mapping between the input at a frequency and port to the output at a specific frequency and port, among the set of modes considered in the analysis.  Explicit formulas for the quantum efficiency with and without dissipation are presented in terms of quantum X-parameters and evaluated for two distinct JTWPA designs for the first time using a commercial circuit simulator. The gain and quantum efficiency results are consistent among ADS, JosephsonCircuits.jl, and WRSPICE. The presented analysis flow is applicable to all commercial harmonic balance solvers and is already implemented in a recent commercial release of ADS that includes a Josephson junction model. Compared to Fourier analysis of a time domain simulation, these methods more rapidly and accurately simulate the multiple interacting signal and idler modes that are required to predict JTWPA quantum efficiency. This analysis more realistically captures the discrete nature of the JTWPA layout (discrete components that can vary along the structure) than analytic continuum or coupled mode theory models.

Detailed layout-dependent parasitic effects, such as bends, can be modeled in the same design environment using EM analysis. The full power of commercial nonlinear circuit simulators is now available to optimize JTWPAs and other devices based on parametric processes (e.g. circulators and kinetic inductance amplifiers) for key quantum applications.

\section{Future work}
Extending the interoperable simulation, modeling, and measurement X-parameter use model \cite{pedro_nonlinear_2018} to include JTWPAs and similar quantum devices with their full characteristics including dissipation and fabrication variation is the ultimate objective. Future work therefore includes calibrated nonlinear X-parameter measurements of real JTWPAs in a cryogenic environment. This will provide a direct comparison of simulations and calibrated nonlinear measured JTWPA performance characteristics, such as gain, quantum efficiency, and out-of-band impedance effects. An enhancement to the existing X-parameter model within Keysight PathWave Advanced Design System (ADS) will enable consumption of simulated and/or measured X-parameters of the JTWPA to be directly consumed as a behavioral model block for the design of qubit readout chains. JTWPA component manufacturers could then provide X-parameter models to quantum chip designers without compromising design IP, and consumers of JTWPA components can specify requirements based on X-parameters and QE derived from them. Users of JTWPAs and other parametric devices can compare prospective technologies in the simulator using the X-parameter behavioral model in a realistic environment including the full readout chain.

\section*{Acknowledgment}
Thanks to Nizar Messaoudi, Chad Gillease and Jason Horn for useful discussions and Keysight management for support. Kaidong Peng and Kevin O'Brien acknowledge funding from the AWS Center for
Quantum Computing and the MIT Center for Quantum Engineering (CQE).

\bibliographystyle{IEEEtran} 

\end{document}